\documentclass[letterpaper,12pt]{article}
\usepackage[english]{babel}
\usepackage{amssymb,amsmath}
\usepackage{endnotes}
\usepackage{graphicx}

\begin{document}

\title{The scientific construction of the world}
\author{Luigi Foschini\\
\small INAF -- Osservatorio Astronomico di Brera, Merate (LC) -- Italy
}

\date{\today}

\maketitle


As soon as I have read the title and the topic of the competition\endnote{{\it FQXi} 2015 Essay Contest, {\tt http://fqxi.org/community/essay/rules}.}, one name and one title flashed in my mind: Wigner, {\it The unreasonable effectiveness of mathematics in the natural sciences} \cite{WIGNER}. After a first look at the competing essays already published\endnote{{\tt http://fqxi.org/community/forum/category/31424}}, I see that I am surely not alone in having had this experience. A consequence of these thoughts was to remind the Wittgenstein's words: according to him, thinking is to operate with signs \cite{WITT}. I am now operating with signs while writing and thinking what to write. The ink lines that I am leaving on the page\endnote{Usually, I prefer writing an essay by hands and then I transcribe it on the computer.} are somehow linked with physical objects. The word ``Wigner'' is connected with a human being lived between 1902 and 1995. The word ``Wittgenstein'' is connected with another human being lived in another epoch, between 1889 and 1951. Already these notes have added new links and new concepts: numbers, dates, years. By looking at these numbers, some readers could remind more events, personal or not, pleasant or not. I could wrote now about the places where these persons lived, generating new connections and ramifications. These dates and places could suggest events to someone or nothing to others. On 1889 April 26, Ludwig Wittgenstein was born in Vienna: this is one fact. However, infinite things happened that day in that place: simply, we selected some in order to define a certain event. Moreover, I was not yet born at that epoch, and therefore I could not be present to see the birth of the Austrian philosopher. Therefore, I have no direct perception, there is no empirical experience of that event for me. Others were present and have written their experience. Thus, the only way for me to know about that event is to read the reports written by others. However, as already written, the words and the numbers used to define that event drag along themselves more meanings. As noted by Carlo Sini\cite{SINI}, even before being empirical matter, a fact is a linguistic expression with a history of meanings and interpretations often forgotten or taken for granted, but making that expression well-acquainted and intelligible. 

The same concepts are applied to science. To make an example, I choose something from my primary research activity, the astrophysics. The example is about the study of the electromagnetic emission of a quasar named 3C~454.3, at coordinates $\alpha = 22^h 53^m 57^s .7$, $\delta = +16^d 08^m 54^s$, and $z = 0.859$ (J2000). We know that the electromagnetic spectrum covers from radio waves to $\gamma$ rays, but there is no instrument able to observe over all the broad frequency or wavelength band. Therefore, it is first necessary to select the frequency range to study. I use the publicly available data of the Large Area Telescope (LAT, \cite{LAT}) onboard the {\it Fermi Gamma-ray Space Telescope} that is operating in the range of $\gamma$ rays with energies between 100~MeV and 100~GeV. This first choice already recalls many more concepts, most of them highly technical: energy, frequency, wavelength, quasar, $\gamma$ rays, MeV, GeV, ... These abstract concepts are all linguistic facts. There is not even some direct perception to link with the words or symbols, because none has ever seen or touched a 100 MeV $\gamma$ photon, and could not even say if it was really a particle or a wave or both. 

Let's go ahead: I connect with the public archive of LAT data\endnote{{\tt http://fermi.gsfc.nasa.gov/cgi-bin/ssc/LAT/LATDataQuery.cgi}.}, and ``download the data''. This simple expression implicitly means that I send an electromagnetic signal to a computer in USA, which in turn answers by sending to my notebook a series of electromagnetic signals changing the state of millions of transistors integrated in the chips of the memory. More electronic devices are changed as consequence: for example, the on/off arrangement of the several pixels of my monitor. I, a human being, interpret these changes of state and light as signs, which in turn are for words or numbers or mathematical symbols (see my discussion in \cite{FOSCHINI1}). A cat or a dog in front of the same monitor would remain unperturbed, or, perhaps, if I would show a video with other cats and dogs, they could start playing without understanding the difference between the animals of the video and themselves. 

I operate on the signs of my monitor to process and to analyse the downloaded data, to extract the light curves\endnote{Although it was not necessary for this essay, I wrote in the following some explanatory notes about the \emph{Fermi}/LAT data analysis. The adopted methods are the same described with more details in \cite{FOSCHINI3} and references therein, but in the present case I worked with {\tt LAT Science Tools v. 9.33.0}, and the {\tt Instrument Response Function P7REP\_SOURCE\_V15}. More light curves of 3C 454.3 can be found in \cite{FOSCHINI2}; in the present work, I have taken as example the exceptional outburst of 2010 November (see \cite{FOSCHINI2}, Fig. 3, bottom right panel).}. These would be the renowned ``experimental data'', which should decide the fate of a theory and tell us how the reality is. However, already the operations performed to date underline that we are just dealing with languages and that the definition of fact or event is just the final product of a series of decisions more or less conscious. I could also add that the ``downloaded'' data, the photons recorded by LAT, are in turn the result of choices: selection of parameters (voltages, currents, temperatures, ...) to define the nominal operations of the instrument, selection criteria and rejection of the cosmic background, and so on. Moreover, as calibrations go ahead, there are updates in the software and the instrument response, which in turn often result in some unexpected change in some sources or spectra or light curves (see, for example, the Sect. 4.2 of \cite{3FGL})\endnote{There is a joke among the researchers saying that the best calibration is obtained when the instrument ended its lifetime, but then none could use it...}. 

\begin{figure*}[t!]
\begin{center}
\includegraphics[angle=270,scale=0.6]{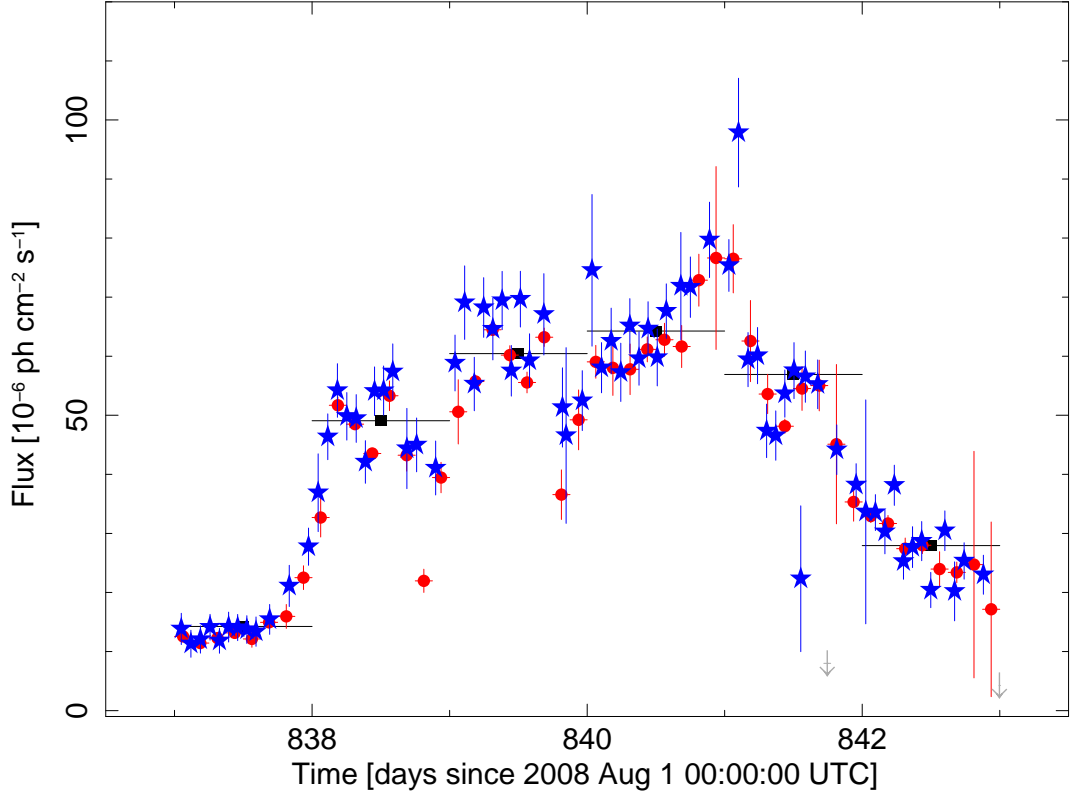}
\caption{Light curves of 3C 454.3 in the period 2010 November 16-22 with different time bin: 1 day (black squares), 3 hours (red circles),  $\sim 1.5$ hours (blue stars). Grey arrows toward down indicate that a statistically insufficient number of photons has been detected, and therefore one sets an upper limit to the flux. The light curve with $\sim 1.5$-hour bin has less points with respect what expected by simply doubling the number of points of the 3-hour bin light curve, because in some cases the quasar was outside the field of view of the instrument. The flux is in units of $10^{-6}$ photons per cm$^2$ per s in the 100~MeV - 100~GeV energy band. See \cite{FOSCHINI2} for more details on this event.}
\label{fig:curva}
\end{center}
\end{figure*}

Anyway, now that I have these photons, I have to decide which are coming from the quasar 3C~454.3 and which are from nearby contaminating sources or from the cosmic background. In these cases, it is usual to adopt a linguistic technical procedure named likelihood \cite{MATTOX}, which in turn indicate the probability that a certain photon is associated to the source under study. To know that a  photon of a certain energy is coming from 3C~454.3 is an important information, but a more useful physical information would be to know the flux of these photons. One individual particle is not sufficient, so that it is necessary to detect a sufficient number of photons, where the term ``sufficient'' has a meaning understandable within the statistical analysis. Therefore, one takes into account the photons collected in certain time intervals (bins). I obtain different light curves depending on the bin width, and thus different information. From the example displayed in Fig.~\ref{fig:curva}, one could see that the source is almost constant at the beginning and the information of the three curves are almost coherent. However, already at the end of the first day, it is possible to see, from the curves with smaller bins, that the source increased its emission. Around the day 841 (November 20), only the light curve with the smallest bin is able to show the sudden peak of emission, while the curves with wider bins the emission is smoothed by the time average. Therefore, what is reality? What is the ``true'' emission of 3C~454.3? 

In addition, this is what I can do with \emph{Fermi}/LAT, the most advanced $\gamma$-ray telescope at hands of mankind, but it still has its technological limits, its sensitivity, its spatial and time resolution, and so on. It is surely more advanced of its predecessors, but its successor will surely have better performance, which will allow us to extract light curves with finer bins, thus giving us different information. So, again, what is reality? What is the ``true'' emission of 3C~454.3? 

Anyway, beyond the boring details of this rather dull presentation -- that are just a negligible part of those that could be made explicit -- I hope it is evident that the so called ``experimental datum'' is just a linguistic fact, with a more or less conscious history of choices and interpretations. One could do infinite experiments and observations, but one select some and not others, one select to measure certain quantities and not others. The empirical fact does not exist by itself, but it is always the result of a linguistic choice. Wigner wrote about an isomorphism about the logic of mathematics and the natural laws \cite{WIGNER}, while Tegmark pushed himself even to say that the universe is a mathematical structure \cite{TEGMARK}. Languages (of any type, spoken, written, and mathematics) are used to define a fact, and are then used to link these facts, the measured quantities. Should one be surprised to find a link between facts defined in a linguistic way and the languages used to define them, both human inventions? It is like to design and to build a house and then be surprised that it stands up... 

Spoken and written languages, jargon, mathematics are all self-organised complex structures emerged from the chaotic interactions of individuals all over the millennia. Only some parts are currently used. We keep memory of other parts, although we think they are obsolete (think about the Ptolemy astronomy, the caloric, the numbers of Ancient Romans,...). Many more parts were lost forever (try thinking about the knowledge lost in the fire of the library of Alexandria). If some linguistic forms were successful to be currently used by most of human beings, this could not even be due to some usefulness parameter: it is known that often scientists were inspired by standards of beauty, which in turn are not something universal, but are changing from person to person. Nevertheless, it remains the fact that all are human constructions, as de Saussure understood \cite{SAUSSURE}, invented not only to communicate, but also to think. 

The wide applicability of mathematics should not be a surprise. It happens also in the spoken language: the words ``be born'' does not refer only to human beings, but to every living being and also to lifeless objects. One could speak about the ``birth'' of a star, isn't it? Just in these days, it happened to me to have at hands the original paper where Weibull defined the homonym distribution \cite{WEIBULL}. It is interesting to note the examples made by the author: yield strength of a Bofors steel; size distribution of fly ash; finer strength of Indian cotton; fatigue life of a St-37 steel; height of British adult males; width of beans. It would be possible to write infinite examples, but it is necessary to note that, with respect to the spoken language, the mathematics tries limiting the semantic field of its signs by means of definitions and the use of different signs, so to avoid intrusion of not-required meanings from everyday life. According to Tegmark \cite{TEGMARK}, the science will reach the complete description of the universe when it will succeed to remove the ``human baggage''. However, it is not possible to remove the human nature from mathematics, because it is a language constructed by human beings\endnote{It would be interesting to see what would happen in the case of contact with alien civilisations different from human beings, but with the same ability to invent languages.}. We could try, but something always overflows, something escapes, consciously or not. This is not bad, because it is the seed of the invention \cite{FOSCHINI4}. 

The fact that our concept of nature, of universe, of world, was a linguistic structure, which in turn was a human invention from which it is not possible to remove the human nature, does not affect its validity and does not influence the reality of things under study. The technological development of the latest four centuries is an evident proof. Simply, it is necessary to abandon a concept of science as research of a unique, absolute, ultimate, fundamental, confident, and soothing Truth. As Niels Bohr wrote, physics is not the study of something given, but rather the development of methods to organise and to measure the human experience \cite{BOHR}. To make observations and experiments that will suggest new events to study, new ways to speak about the nature. In the end, we continue speaking even though we know that there are in the language some undecidable phrases like ``I am lying''.

\theendnotes

\end{document}